# STATISTICAL SHAPE MODELLING OF THE NASOSINUSAL ANATOMY

Michele Bertolini[1], Silvia Tonghini[1], Marco Rossoni[1], Marina Carulli[1], Susanne Weise[2], Christian Jan Baldus[2], Giorgio Colombo[1], Monica Bordegoni[1]

[1]Department of Mechanical Engineering, Politecnico di Milano, Milan, Italy
[2]Technischen Universität Dresden, Dresden, Germany

**ABSTRACT**

*The human nose exhibits a huge variation in shape among individuals. All these variants alter the airflow through the nasal cavity and can impact how we smell odors. To acquire a better understanding of physiological and pathological functioning, it is important to study the effects of these modifications. SSM, or Statistical Shape Modelling, is a widely used methodology for considering morphological differences within a population.*

*In the literature, only a few studies analyze nasal anatomy in its entirety, including paranasal sinuses, whose segmentation is particularly challenging due to their complexity and variability. This work aims at creating a highly accurate SSM for the nasosinusal complex.*

*A traditional, greyscale thresholding-based segmentation approach was chosen, relying on Mimics software (Leuven, Belgium). 40 computed tomography (CT) datasets were considered. After segmentation, post-processing was performed to obtain watertight meshes using 3-Matic software (Leuven, Belgium), which was also used for the SSM generation.*

*The generated model shows all relevant landmarks that typically characterize a nasosinual complex. 32 different modes are needed to explain 95% of the total variance.*

*As a preliminary output, this study showed the feasibility and consistency of a statistical model of the nasosinusal anatomy. However, additional studies involving a larger sample size and a more robust validation process could be conducted.*

## 1. INTRODUCTION

The human nose is an important and complex organ that performs several functions, such as respiration, air humidification and filtration, and olfaction. It exhibits large variations in shape among individuals due to for example a different bone structure [1]. This anatomical variability critically affects nasal airflow and thus nasal function [2]. For this reason, it is important to carefully investigate it.

The sense of smell significantly impacts people's lives, and any impairments on it affect individuals' daily functioning. The total or partial loss of smell affects up to 24% of the global population [3]. Although erroneously considered a "minor" impairment, it profoundly impacts the quality of life [3]. Despite this, there hasn't been progress in developing devices to restore the sense of smell due to insufficient scientific understanding of connecting artificial systems with human olfaction. The ROSE project, funded under the Horizon 2020 FET-Open program, aims to address this gap by generating new scientific insights that will culminate in creating the Digital Olfaction Module (DOM). This breakthrough technology will allow patients to perceive their olfactory surroundings, enhancing their overall sensory experience.

To achieve this result, the ROSE project combines interdisciplinary competencies and skills related to the study of the human olfaction system to develop a proof of concept combining miniaturized odor sensors and stimulation arrays that will be evaluated in patients with smell disorders. The project crucially involves designing and developing prototypes of bespoke miniaturized artificial noses to integrate and accurately position the smell sensor around/inside the nostril. Specifically, the goal is to design a device that properly places the sensor according to the morphology of the subject's nasal cavity. For this reason, it becomes very important to accurately know and describe it, with its possible anatomical variations.

The typical imaging technique to evaluate the nose is computed tomography (CT), which represents the golden standard in the diagnostics of nasal cavity diseases [4]. Starting from CT data, through a segmentation process, it is possible to extract a 3D digital anatomical model of the nose. Obtained patient-specific model can be exploited to get a better understanding of normal and pathological functioning, even when coupled with additive manufacturing technology, to create physical prototypes for experimental testing [5] or with numerical simulations, to study the inner airflow [6]. However, to account for anatomical variability in a robust and comprehensive way, a different approach is needed.

As already mentioned, the huge anatomical variability is a crucial aspect to be considered in this design process of these miniaturized artificial noses, and for this reason a tool able to account for and describe it in an effective and comprehensive way, but also to generate new, unseen synthetic geometries, is needed.

Statistical Shape Modelling (SSM) is a powerful mathematical tool that typically exploits Principal Component Analysis (PCA) to generate a synthetic and comprehensive representation of complex shapes. In the medical field, it is nowadays a widely used tool to describe the local morphological variability in a population, in a continuous way. The resulting principal components (*modes*) encode the most dominant features that determine the shape variation, in a weighted descending order. The creation of an SSM of the nose opens many opportunities for further analysis [7],[8],[9]: it can be useful to study the influence of morphology on different flow characteristics of the nasal flow, e.g. nasal resistance, heat transfer, and flow humidity, but also for clinical studies related to particle deposition, drug delivery, and disease diagnosis and treatment, including medical devices design.

Some studies in the literature have tackled the problem of nasal SSM generation. Brüning et al. [10], for example, generated a standardized geometry of the healthy nasal cavity based on CTs from 25 subjects, to perform fluid dynamics simulations and evaluate if the airflow in this averaged geometry can be representative of the healthy nasal airflow. Similarly, SSM of the human nose based on a combination of CT and CBCT images of 46 patients, with very different nasal or sinus related complaints, was obtained by Keustermans et al. [7], with a technique based on cylindrical parametrization. The generated model was exploited for two application examples, namely nasal shape evaluation in function of age and gender, and a morphometric analysis of different anatomical regions. Eventually, in [11] a 100 pathological patients clinical training set was built and used to investigate the effects of shape on nasal heating function.

Even if analyzed studies produced relevant representations of the nasal cavity and its morphological variations, none of them considers the entire complex, including paranasal sinuses, namely ethmoid, frontal and maxillary sinuses. Paranasal sinuses are essential components as they fulfil various functions: they protect the organism, mostly by humidifying the inhaled air and assisting the immune response of the respiratory system. Moreover, the sinuses are air-filled spaces that significantly decrease the weight of the head and affect the resonance of the human voice. Segmenting the whole geometry is important for diagnosis, treatment, and research, but it is also very challenging due to its complex and variable anatomy. For this reason, they are often excluded from the 3D model generation of the nasal cavity [12]. Furthermore, applying a challenging approach like SSM to such a highly complex anatomy, is something new within the field, to our knowledge.

The goal of this work is to generate and validate a robust SSM of the nasosinusal complex, also identifying the primary patterns of shape variation. Starting from relevant CT datasets, 3D anatomical models are segmented and post-processed for subsequent SSM generation. Then, a preliminary evaluation of the obtained model was performed.

## 2. MATERIALS AND METHODS

In this retrospective investigation, anonymized clinical CT scans from about 60 patients were considered. Only adults who exhibited no pathology in the nasal cavity were included.

Among them, 40 were considered qualitatively sufficient from a technical point of view to be used in the model creation. Indeed, the quality of CT scans is crucial in the process, and only images with sufficient resolution and Signal-to-Noise Ratio (SNR) are suitable. In particular, the study included 22 males and 18 females aged between 19 and 84. The dataset information considered is listed in Table 1.

A traditional, greyscale thresholding-based segmentation approach for the whole nasal cavity geometry has been chosen, relying on Mimics software (Leuven, Belgium). The lower threshold value remained constant, corresponding to the minimum Hounsfield Unit [HU] value (-1024), while the upper one was arbitrarily adjusted according to the specific dataset, to avoid under- or oversegmentations.

**TABLE 1:** LIST OF INCLUDED DATASETS, WITH PATIENTS' AGE AND SEX, FOR SSM GENERATION.

| # | age | sex |
|---|-----|-----|
| *1* | 64 | M |
| *2* | 37 | F |
| *3* | 58 | F |
| *4* | 52 | M |
| *5* | 28 | M |
| *6* | 39 | M |
| *7* | 29 | F |
| *8* | 44 | M |
| *9* | 29 | F |
| *10* | 42 | M |
| *11* | 36 | M |
| *12* | 20 | M |
| *13* | 19 | M |
| *14* | 29 | M |
| *15* | 26 | M |
| *16* | 32 | F |
| *17* | 48 | F |
| *18* | 34 | M |
| *19* | 42 | F |
| *20* | 28 | F |
| *21* | 77 | F |
| *22* | 40 | F |
| *23* | 26 | M |
| *24* | 44 | F |
| *25* | 45 | M |
| *26* | 63 | F |
| *27* | 44 | F |
| *28* | 72 | F |
| *29* | 42 | F |
| *30* | 59 | F |
| *31* | 69 | M |
| *32* | 52 | M |
| *33* | 64 | M |
| *34* | 55 | M |
| *35* | 54 | M |
| *36* | 84 | F |
| *37* | 33 | M |
| *38* | 40 | M |
| *39* | 62 | F |
| *40* | 22 | M |

Each anatomy was divided into four masks, containing the cavity, the ethmoid, the frontal and the maxillary sinuses, respectively. Being manual segmentation a time-consuming process, highly dependent on the operator, each of these steps was done by the same operator for all 40 CT scans to avoid potential inter-operator variability introduced in the process. An example of segmentation result is reported in Fig. 1.

After segmentation, post-processing was systematically performed using 3-Matic software (Leuven, Belgium). A sequence of *Wrap*, *Adaptive Remesh*, and *Smooth* was applied to obtain coherent, watertight geometries. The *Wrap* operation creates a wrapping surface of the selected entities and this is helpful to filter small inclusions and close small holes resulting from the segmentation process. Moreover, inverted normals, bad contours, small gaps and unconnected triangles are also fixed by the process, making the 3D model watertight. The *Adaptive Remesh* rebuilds the topology of the mesh while preserving its geometry. Finally, a smoothing filter has been applied to decrease "noise" in the mesh. The smoothing filter selected in this case is a curvature-based approach which was considered the most appropriate to preserve the shape of the mesh. The last step before creating the SSM has been a boolean union between the nasal cavity and sinuses: this step is necessary to consider the whole anatomy as a unique entity in the following steps.

For SSM generation, a module available in 3-Matic software was exploited. First of all, one of the models was randomly chosen to become the reference mesh and a rough registration of the remaining 39 cavities was performed. The rough registration has been achieved by applying translations and rotations to each model to get a first-tentative alignment with the reference one. Then, the fine registration has been performed through the Iterative Closest Point (ICP) algorithm, exposed by 3-Matic through the global registration function. This algorithm minimizes the distance between the reference part and a moving entity. The procedure has been applied systematically between the reference mesh and all the other 39 cavities. For each pair, the global registration was applied more than once: being the ICP algorithm strongly non-linear, these repetitions are useful to ensure the convergence of the residual alignment error between the two meshes. A uniform remeshing was then applied to the chosen reference, setting a 2 mm maximum edge length. This re-meshed cavity served as the "master" mesh. Then, a registration process involving scaling was iteratively performed between each model, designated as the "fixed entity", and the master mesh, designated as the "moving entity". This process aimed to scale and globally align the "master" mesh with each one of the 39 input models as closely as possible. At this point, the transformation matrices mapping (i.e., morphing) the reference model into the single-patient cavity were known: applying the transformation matrices to the master it was possible to generate the model of the corresponding patient cavity. Similarly, a *Warp* operation was applied between each model and the master, to create point correspondence. This operation produced, for each model, a new one that contained the same number and position of triangles in comparison to the "master" (Fig. 2). Eventually, it was possible to combine them and generate the SSM.

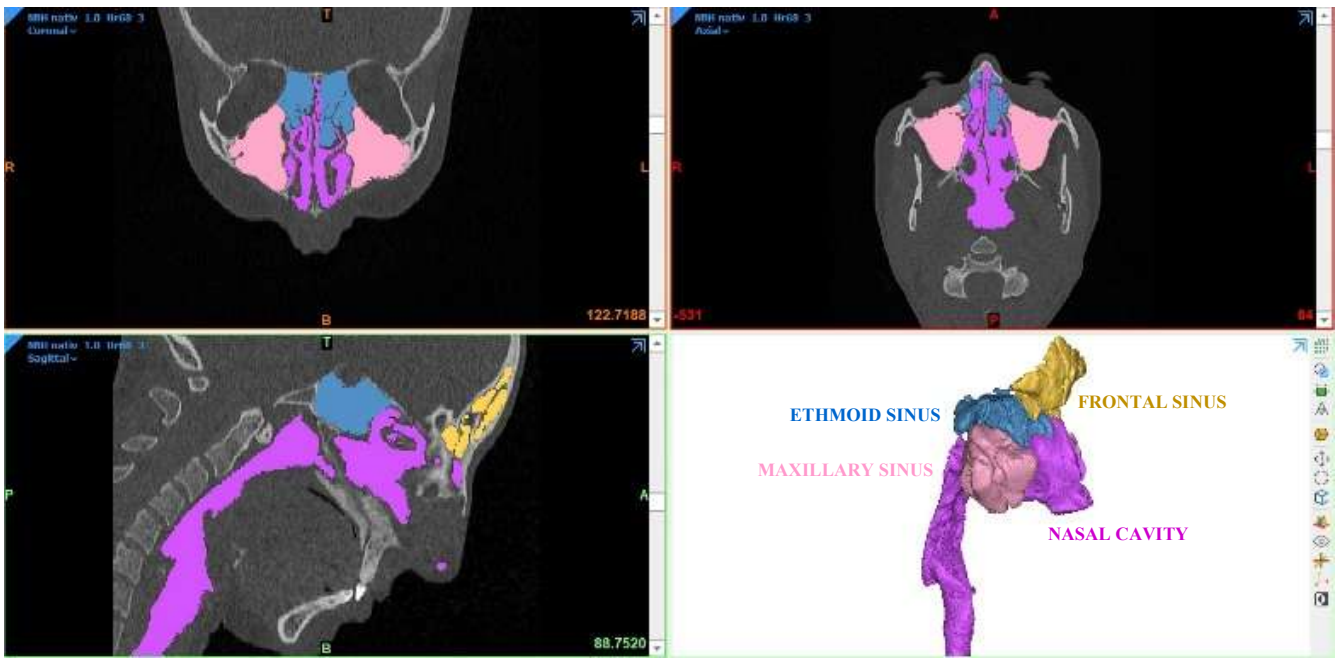

**FIGURE 1:** EXAMPLE OF THE SEGMENTATION PROCESS OF A CT DATASET IN MIMICS. DIFFERENT MASKS ARE SHOWN IN DIFFERENT COLOURS.

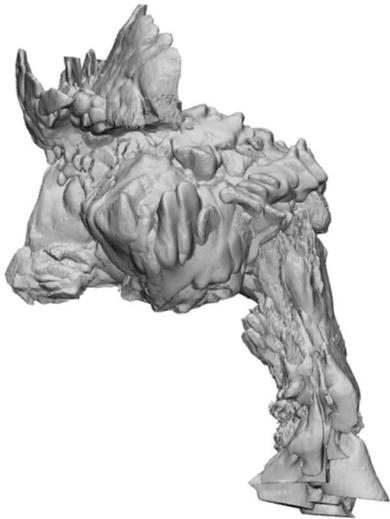

**FIGURE 2:** THE CONSIDERED 40 MODELS, AFTER ALIGNMENT AND REGISTRATION.

This is the ground on top of which the statistic was built. Specifically, a principal component analysis (PCA) was performed to determine the principal modes of shape variation on the set of samples previously described. PCA ranks the modes of shape variation by their explained variance. In the case of this work, a cut-off threshold has been decided according to the cumulative variance, as later discussed.

## 3. RESULTS AND DISCUSSION

The obtained SSM is reported in Fig. 3. The geometry features all relevant landmarks that typically characterize a nasosinual complex.

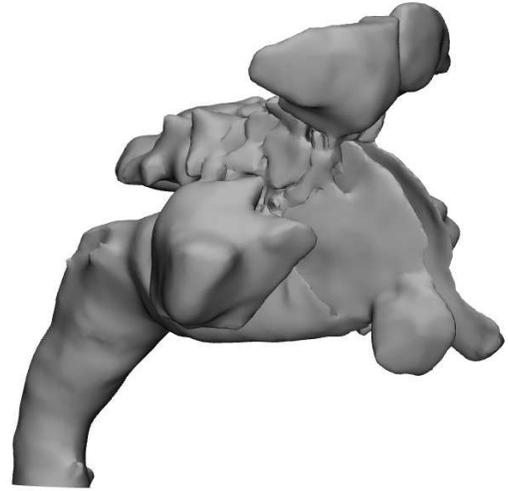

**FIGURE 3:** THE GENERATED STATISTICAL MODEL OF THE NASOSINUSAL COMPLEX.

A usual metric to evaluate how much the SSM describes the variability of the population is the *compactness* of the model. This metric measures how many principal components are needed to describe a certain amount of variability. A model can be defined as "compact" when it requires few parameters to define an instance [13]. Specifically, an SSM is driven by the amount of standard deviation (SD, or σ) associated with each mode of variation. Analogously to a parametric CAD model, the independent parameters that a user can play with to modify the geometry are the SDs associated with each mode.

The shape variance explained by the shape modes derived from PCA was extracted. Fig. 4 shows the cumulative variance of the first 32 components. Modes are sorted in a decreasing way, so that the first one explains most of the variance, accounting for the 12.8% in our case, while the relative weight of the following ones is lower and lower. The first 32 components reported in the chart are able to account for 95% of cumulative variance that can be observed in the population. In other words, the number of independent parameters that can adequately describe the variation of nasosinual complex in our case is 32.

5    © 2024 by ASME

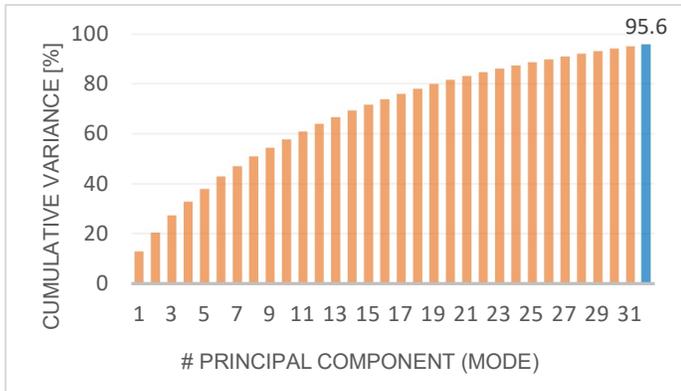

**FIGURE 4:** CUMULATIVE VARIANCE OF THE FIRST 32 MODES.

The next step when deriving an SSM is usually to correlate each mode to a specific (geometrical) variation in the 3D model. This is usually done by changing the value of SD associated with one mode, keeping the other unchanged, and looking at how it affects the overall shape. The range of variation is usually set as ±3SD. It must be noted that, despite patterns could be found, the modes will be always correlated. In our case, in particular, due to the complexity of the anatomy, it was quite difficult to find correspondences with specific patterns of shape variation.

As typical in SSM, and so in the case here discussed, the first mode variation is related to an overall variation in the scale of the anatomy: in other words, a modification in size in all three components of the scale transformation matrix. Trends of variations in the scale factor along x, y and z as well as the global volume by modifying the standard deviation of *mode 1* are reported in Fig. 5 and Fig. 6, respectively. In particular, figures show the percentage variation with respect to the scale factor and the volume at the reference model (i.e. the one with a $0\sigma$ variation). As shown, by increasing the absolute value of the coefficient up to 2.5 SD, the model tends to shrink, reaching about half of the volume corresponding to -2.5 SD. The trend is quite similar for each of the components.

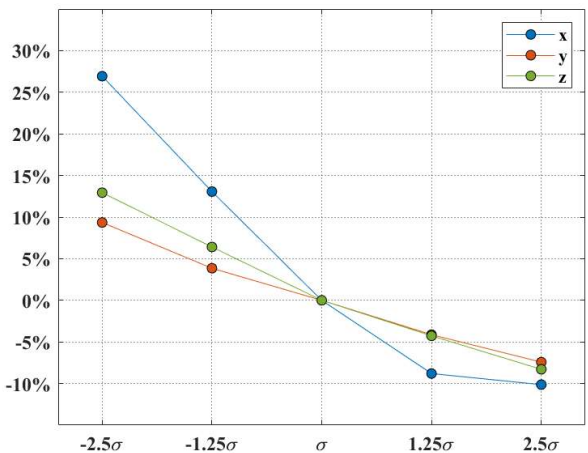

**FIGURE 5:** VARIATION TREND FOR X, Y AND Z SCALE FACTOR OF THE SSM MODEL WHILE CHANGING THE SD ASSOCIATED WITH MODE 1.

Moving to the modes greater than one, the pattern identification becomes too difficult to isolate: several geometrical modifications get strictly interconnected, making the correlation between the SD associated with each single mode to a certain geometrical parameter not possible.

Moreover, moving towards a coefficient value equal to ±3 SD most of the time results in significant distortions and artifacts in the geometry. This could be explained by the huge anatomical variability intercurrent among models considered for the SSM generation, together with the intrinsic complexity of an anatomy where paranasal sinuses are considered, too.

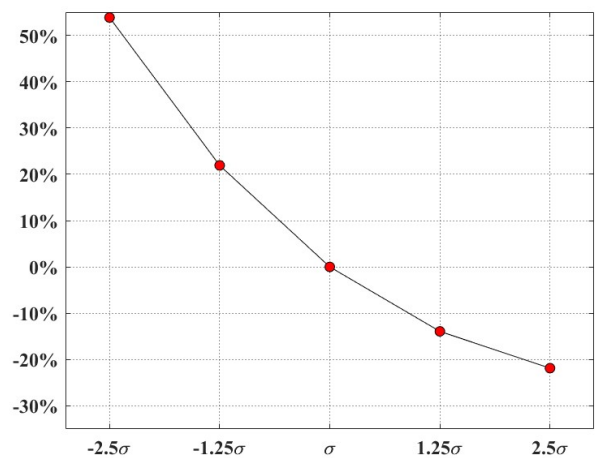

**FIGURE 6:** VARIATION TREND FOR THE VOLUME OF THE SSM MODEL WHILE CHANGING THE SD ASSOCIATED WITH MODE 1.

## 4. CONCLUSION

The goal of this work was to generate and validate a robust SSM of the nasosinusal complex, also identifying the primary patterns of shape variation. Starting from relevant CT datasets, 3D anatomical models have been segmented and post-processed for subsequent SSM generation. A preliminary evaluation of the obtained model was performed, in terms of its compactness and ability to identify variation patterns corresponding to the different modes.

Compared to other works in the literature, the present study is relevant in two ways. First of all, to our knowledge, none of the previous works considered the entire nasosinusal complex, including paranasal sinuses. Secondly, from our study we could notice how, even with an increased anatomical complexity, the compactness of the SSM is similar to the one of previous analogous studies.

 

Table 2 reports sample size and number of modes needed to account for 95% of variance for two similar works, compared with the SSM here generated. In both of them, segmentation and SSM generation were limited to the nasal cavity. As it can be noticed, the number of modes needed to reach a 95%-variance is in line (even lower) with them. Hence, the approach used to generate the proposed SSM seems to be feasible and robust, even when paranasal sinuses are taken into account.

**TABLE 2:** SAMPLE SIZE AND NUMBER OF MODES IN DIFFERENT STUDIES.

| Reference | Sample Size [patients] | Number of modes (95% SD) |
|---|---|---|
| [7] | 46 | 46 |
| [10] | 25 | 33 |
| Ours | 40 | 32 |

Further investigations should be conducted by extending the population and a more systematic validation procedure could include also the assessment of *generalization* of the SSM, namely how well a random anatomy can be described by the model, and *specificity*, measuring the soundness of new shape instances randomly generated by the shape model [11].

On top of this, future works will also include a more "practical" approach to evaluate the real effectiveness of the SSM. In the context of the ROSE project, a DOM is being developed. This device, whose primary function is to act as a prosthesis for people affected by anosmia, is composed of two main devices: a stimulator, that sends signal to the brain, and a detection module which oversees the detection of the odors in the surrounding environment. Optimizing the position of the detection module is of primary importance for the project. However, as discussed in this paper, the nasosinusal complex shows great variability. So, an SSM that is compact and provides a good representation of the population would allow experiments and simulations to be conducted easily, faster and with greater confidence in results. Hence, the next steps will be to perform experiments on 3D printed replicas or numerical simulations (relying on computational fluid dynamics) using the SSM derived in this work and to compare the result with the patient-specific models ones [14][15].

## ACKNOWLEDGEMENTS

The ROSE Project has received funding from the European Union's Horizon 2020 research and innovation program under grant agreement No 964529 (Pathfinder ROSE project).